\documentclass[pra,aps,showpacs,floatfix,twocolumn,nofootinbib,superscriptaddress]{revtex4-1}
\usepackage{amsmath}
\usepackage{amsfonts}
\usepackage{amssymb}
\usepackage{revsymb}
\usepackage{graphicx}
\usepackage{mhchem}
\usepackage{siunitx}
\usepackage{multirow}

\def\ket#1{|\,#1 \,\rangle}

\def\vop#1{\mathbf{\hat{#1}}}

\begin{document}
\title{Magic wavelength of the $^{138}$Ba$^+$ $6s\;{}^2S_{1/2} - 5d\;{}^2D_{5/2}$ clock transition }
\author{S. R. Chanu}
\affiliation{Centre for Quantum Technologies, National University of Singapore, 3 Science Drive 2, 117543 Singapore}
\author{V. P. W. Koh}
\affiliation{Department of Physics, National University of Singapore, 2 Science Drive 3, 117551 Singapore}
\author{K. J. Arnold}
\affiliation{Centre for Quantum Technologies, National University of Singapore, 3 Science Drive 2, 117543 Singapore}
\author{R. Kaewuam}
\affiliation{Centre for Quantum Technologies, National University of Singapore, 3 Science Drive 2, 117543 Singapore}
\author{T. R. Tan}
\affiliation{Centre for Quantum Technologies, National University of Singapore, 3 Science Drive 2, 117543 Singapore}
\affiliation{Department of Physics, National University of Singapore, 2 Science Drive 3, 117551 Singapore}
\author{Zhiqiang Zhang}
\affiliation{Centre for Quantum Technologies, National University of Singapore, 3 Science Drive 2, 117543 Singapore}
\author{M. S. Safronova}
\affiliation{Department of Physics and Astronomy, University of Delaware, Newark, Delaware 19716, USA}
\affiliation{Joint Quantum Institute, National Institute of Standards and Technology and the University of Maryland,
College Park, Maryland, 20742}
\author{M. D. Barrett}
\email{phybmd@nus.edu.sg}
\affiliation{Centre for Quantum Technologies, National University of Singapore, 3 Science Drive 2, 117543 Singapore}
\affiliation{Department of Physics, National University of Singapore, 2 Science Drive 3, 117551 Singapore}
\begin{abstract}
The zero crossing of the dynamic differential scalar polarizability of the $S_{1/2}-D_{5/2}$ clock transition in $^{138}$Ba$^+$ has been determined to be $459.1614(28)\,\mathrm{THz}$.  Together with previously determined matrix elements and branching ratios, this tightly constrains the dynamic differential scalar polarizability of the clock transition over a large wavelength range ($\gtrsim 700\,\mathrm{nm}$).  In particular it allows an estimate of the blackbody radiation shift of the clock transition at room temperature.
\end{abstract}
\maketitle
Singly-ionized barium has been well studied over the years with a wide range of precision measurements \cite{marx1998precise,knoll1996experimental,hoffman2013radio,woods2010dipole,sherman2008measurement,kurz2008measurement,munshi2015precision,dutta2016exacting} that have provided valuable benchmark comparisons for theory \cite{guet1991relativistic,dzuba2001calculations,iskrenova2008theoretical,safronova2010relativistic,sahoo2007theoretical,gopakumar2002electric}.  It was recently proposed that some of these measurements, specifically high accuracy measurements of transition matrix elements and branching ratios, could be used to construct an accurate representation of the dynamic differential scalar polarizability, $\Delta\alpha_0(\omega)$, of the $S_{1/2}-D_{5/2}$ clock transition over a large wavelength range \cite{barrett2019polarizability}.  Crucial to that proposal was a determination of a zero crossing $\Delta\alpha_0(\omega_0)=0$ near $653\,\mathrm{nm}$, which bounds significant contributions to $\Delta\alpha_0(\omega)$ from the ultraviolet (uv) spectrum.  Here we determine $\omega_0$ with an inaccuracy of a few GHz.  

Together with measurements of the $P_{3/2}$ branching ratios \cite{arnold2020branching} and reduced matrix elements \cite{woods2010dipole}, our result allows us to construct a model of $\Delta\alpha_0(\omega)$ for the $S_{1/2}-D_{5/2}$ clock transition that is accurate at the sub-\% level for wavelengths above 700\,nm.  This will allow accurate measurement of differential polarizabilities in other ions through comparison with Ba$^+$.  In particular,  it will enable more than an order of magnitude improvement in the polarizability assessments for the $^{176}$Lu$^+$ clock transitions \cite{arnold2018blackbody}.

To find $\omega_0$, a linearly polarized laser beam near 653\,nm is focussed onto the ion to induce an ac Stark shift of the clock transition and the shift measured as a function of the laser frequency $\omega$. The ac Stark shift, $\delta_s(M_J)$, induced by the 653-nm laser is given by
\begin{multline}
\delta_s(M_J)=-\frac{1}{2}\Delta\alpha_0(\omega)\langle E^2\rangle\\
-\frac{1}{4}\alpha_2(\omega)\frac{3 M_J^2-J(J+1)}{J(2J-1)}\left(3\cos^2\theta-1\right)\langle E^2\rangle,
\end{multline}
where $M_J$ denotes the applicable eigenstate of $D_{5/2}$, $\alpha_2(\omega)$ is the dynamic tensor polarizability of $D_{5/2}$, and $\theta$ is angle between the 653-nm laser polarization and the quantization axis.  The tensor component can be eliminated by determining the average ac Stark shift
\begin{align}
\delta_0(\omega)&=\frac{1}{3}\left[\delta_s(1/2)+\delta_s(3/2)+\delta_s(5/2)\right]\\
&=-\frac{1}{2} \Delta\alpha_0(\omega)\langle E^2\rangle
\end{align}
Hence, with the laser intensity fixed, $\delta_0(\omega)$ is directly proportional to $\Delta\alpha_0(\omega)$, which is approximately linear in a neighbourhood of $\omega_0$.  Consequently, $\omega_0$ can be determined from a linear fit to measurements of $\delta_0(\omega)$ as a function of $\omega$ with an accuracy limited by the projection noise of the measurements, and a small nonlinearity in $\Delta\alpha_0(\omega)$, which can be estimated from theory.   

Although the laser power is actively stabilized outside the experiment chamber, etaloning effects and uncalibrated frequency response of the detector can give rise to a frequency dependence of the resulting laser intensity at the ion.  In addition, pointing stability of the laser can also degrade the accuracy of the ac Stark shift measured at a single frequency.  To compensate these effects, one can make use of the weighted average
\begin{align}
\delta_2(\omega)&=\frac{25}{42}\left(\delta_s(5/2)-\frac{1}{5}\delta_s(3/2)-\frac{4}{5}\delta_s(1/2)\right)\\
&=-\frac{1}{4}\alpha_2(\omega)\left(3\cos^2\theta-1\right)\langle E^2\rangle
\end{align}
which is a measure of the tensor polarizability.  The ratio $\delta_0(\omega)/\delta_2(\omega)$ is then independent of slow variations in the intensity but has the same zero crossing.  In addition, setting $\theta\approx \pi/2$ minimizes the influence of magnetic field pointing stability, which would compromise the stability of $\delta_2(\omega)$.

Measurements are carried out in a linear Paul trap with axial end-caps, the details of which have been given elsewhere \cite{arnold2018blackbody,kaewuam2018laser}.  The rf potential at a frequency near $\Omega_\mathrm{rf}=2\pi\times 20.7\,\mathrm{MHz}$ is delivered via a quarter-wave helical resonator. Together with static end-cap and bias potentials, the measured trap frequencies of a single $^{138}$Ba$^+$ are $\sim (810,925,245)\,\mathrm{kHz}$, with the lowest frequency along the trap axis.

\begin{figure*}
\begin{center}
  \includegraphics[width=\linewidth]{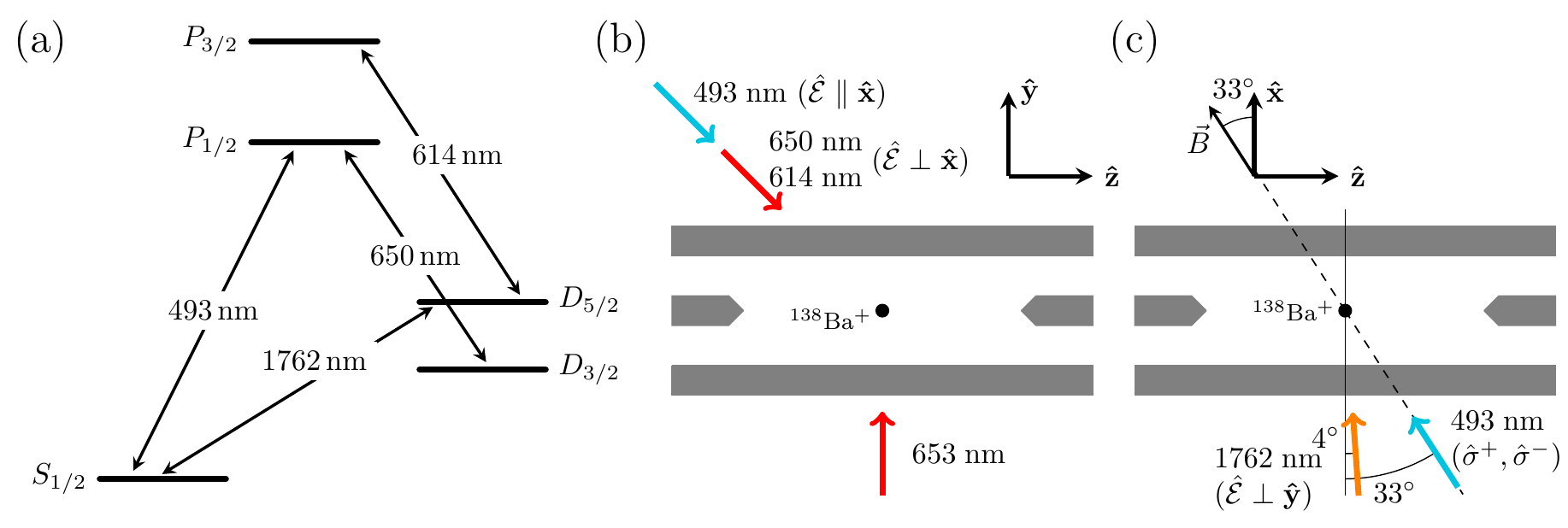}
  \caption{(a) Level structure showing the Doppler cooling and detection transitions at 493 and 650\,nm, the clock transition at 1762\,nm, and the repumping transition at 614\,nm. (b) and (c) Laser configurations used in the experiment. The $\sigma^+$ ($\sigma^-$) beams at 493\,nm facilitate optical pumping into $\ket{S_{1/2},+1/2}$ ($\ket{S_{1/2},-1/2}$).  The 653-nm is used to induce an ac Stark shift and is linearly polarized orthogonal to the magnetic field. Orientation and polarization of the 1762-nm clock laser permits efficient coupling to the $\ket{S_{1/2},\pm 1/2}$ to $\ket{D_{5/2},\pm M_J}$ transitions for $M_J=1/2,3/2,$ and $5/2$.}
  \label{Fig1}
\end{center}
\end{figure*}

The relevant level structure of $^{138}$Ba$^+$ is illustrated in Fig.~\ref{Fig1}(a). Doppler cooling is provided by driving the 493- and 650-nm transitions, with light scattered at 650\,nm collected onto a single photon counting module (SPCM) for detection.  The $D_{5/2}$ level is populated by driving the $S_{1/2}-D_{5/2}$ clock transition at 1762\,nm and depopulated by driving the $D_{5/2}-P_{3/2}$ transition at 614\,nm.  State preparation into the $m=\pm1/2$ states of the $S_{1/2}$ levels is facilitated by two additional 493-nm beams that are $\sigma^\pm$ polarized.

Laser configurations are illustrated in Fig.~\ref{Fig1}(b) and (c). Cooling and repumping light at 493, 614, and 650\,nm are all collinear and propagate at 45 degrees to the $\vop{z}$-axis.  All three beams are linearly polarized with the  650-nm and 614-nm lasers polarized orthogonal to the 493-nm laser, which is polarized along the imaging axis ($\hat{\mathbf{x}}$).  The magnetic field lies in the horizontal ($xz$) plane at an angle of approximately 33 degrees to the $\hat{\mathbf{x}}$-axis with a magnitude of $\sim0.2\,\mathrm{mT}$, which is sufficient to lift the Zeeman degeneracy and prevent dark states when driving the $D_{3/2}-P_{1/2}$ and $D_{5/2}-P_{3/2}$ transitions.  The clock laser at 1762\,nm propagates almost co-linear with the imaging axis and is linearly polarized in the horizontal plane, which permits efficient coupling to the $\ket{S_{1/2},\pm 1/2}$ to $\ket{D_{5/2},\pm M_J}$ transitions for $M_J=1/2,3/2,$ and $5/2$.  A Stark-shifting laser near 653\,nm propagates along $\vop{y}$ and is linearly polarized orthogonal to the magnetic field.  All lasers are switched by acousto-optic modulators (AOMs).

The 653-nm laser is provided by a tunable extended-cavity-diode laser (ECDL) together with an injection locked slave to boost the power.  The laser is locked to a wavemeter with 10\,MHz accuracy (High finesse WS8-10), which is routinely calibrated to a caesium-locked 852-nm diode laser.  Control of the radio frequency (rf) power driving a switching AOM is used to stabilize the beam power, which is sampled by a beam pick-off before the vacuum chamber. 

The 1762-nm clock laser is an ECDL, which is phase locked to an optical frequency comb (OFC).  The short term ($<10\,$s) stability of the OFC is derived from a $\sim 1\,\mathrm{Hz}$ linewidth laser at 848 nm which is referenced to a 10\,cm long ultra-low expansion (ULE) cavity with finesse of $\sim4\times10^5$. For longer times ($\gtrsim 10\,\mathrm{s}$), the OFC is steered to an active hydrogen maser (HM) reference.  The 1762-nm laser is switched with an AOM, but frequency shifting is achieved via a wideband electro-optic modulator (EOM) with the lower sideband driving the transition of interest.  Control of the rf power driving the EOM enables equal $\pi$-times for each clock transition. 

A typical experiment consists of four steps: $1\,\mathrm{ms}$ of Doppler cooling, optical pumping for $50\,\mathrm{\mu s}$ to either $\ket{S_{1/2},m=\pm 1/2}\equiv\ket{S,\pm}$, clock interrogation (Rabi) to $\ket{D_{5/2},\pm M_J}$ for $1\,\mathrm{ms}$, and finally detection for $1\,\mathrm{ms}$.  The initial Doppler cooling step includes the 614-nm beam to facilitate repumping from the upper clock state.  When including the 653-nm beam to Stark shift the transition, the beam is switched on $200\,\mathrm{\mu s}$ before the clock laser, to ensure the beam power has stabilized.  

The clock transition is realized by steering the clock laser to the average of the Zeeman pair $\ket{S_{1/2},\pm 1/2}\leftrightarrow \ket{D_{5/2},\pm |M_J|}$ to cancel linear Zeeman shifts of both levels. Using an interleaved servo in which the optical transition is alternately interrogated both with and without the 653-nm laser, the shift is determined from the difference frequency in the two configurations.  In a single clock cycle, all three Zeeman pairs are measured: each of the six transitions $\ket{S_{1/2},\pm1/2} - \ket{D_{5/2},\pm M_J}$ for $M_J=1/2, 3/2,$ and $5/2$ is interrogated both sides of each transition, both with and without the 653-nm Stark shift beam.  All 24 measurements are repeated $N=50$ times.  From these measurements error signals are derived to track the center and splitting for each Zeeman pair $\ket{S_{1/2},\pm1/2} - \ket{D_{5/2},\pm M_J}$, both with and without the Stark shift.  The ratio $\delta_0(\omega)/\delta_2(\omega)$ is inferred at each step from the appropriate frequency differences.  The servo is first run for several clock cycles, to ensure each shifted and unshifted Zeeman pair is being tracked, and $\delta_0(\omega)/\delta_2(\omega)$ is inferred from a further 1000 cycles.  

\begin{figure}
\begin{center}
  \includegraphics[width=0.9\linewidth]{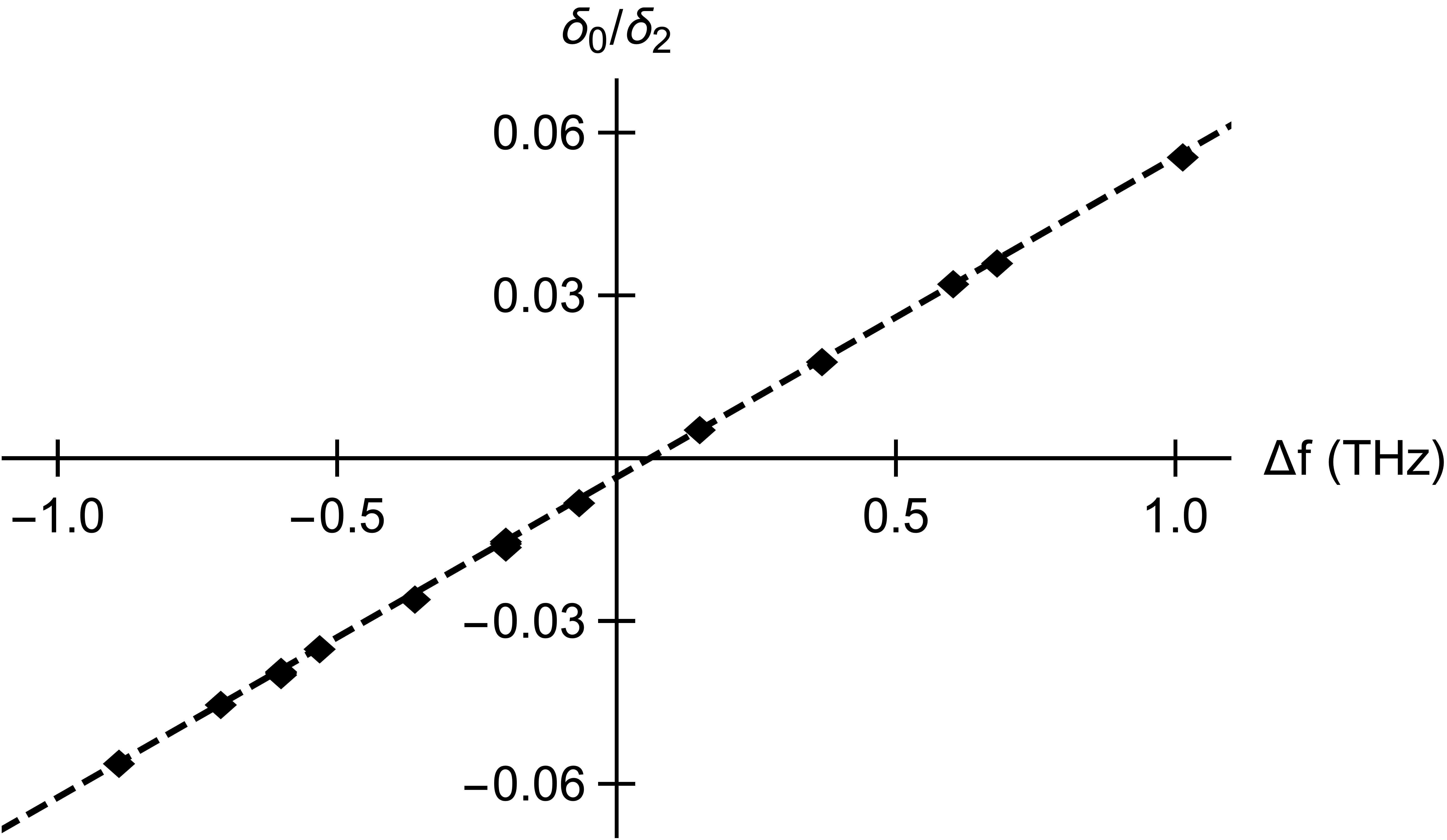}
  \caption{Ratio of scalar to tensor shifts as a function of frequency difference relative to 459.1\,THz. Error bars are smaller than the points and have been omitted.  Measurements at $\Delta f\approx-0.2,-0.6\,\mathrm{THz}$ were measured 3 times giving 16 measurements in total. Dashed line is from a $\chi^2$-fit.}
  \label{Fig2}
\end{center}
\end{figure}

Measurements of $\delta_0(\omega)/\delta_2(\omega)$ were made over a frequency range of approximately $\pm1\,\mathrm{THz}$ and are plotted in Fig.~\ref{Fig2}.  Measurements at $\Delta f\approx-0.2,-0.6\,\mathrm{THz}$ were measured 3 times giving 16 measurements in total.  The fit is a linear regression using the projection noise expected from the 1000 measurements as uncertainties.  The fit has $\chi^2_\nu=1.15$ and an estimated zero crossing of $\Delta f=59.2(1.8)\,\mathrm{GHz}$.  However this does not account for bias arising from a linear fit to $\delta_0(\omega)/\delta_2(\omega)$.

To estimate the bias, we use theory to calculate $\delta_0(\omega)/\delta_2(\omega)$ at the frequencies used in the measurements, fit the resulting points to a straight line, and compare the interpolated zero crossing to that calculated from theory.  This procedure indicates that a straight line fit would underestimate the zero crossing by $\approx 2.2\,\mathrm{GHz}$.  This does not change significantly if one offsets the frequencies used in the calculation by the small difference between the experimentally measured and calculated zero crossing.  Consequently we apply $2.2\,\mathrm{GHz}$ as a correction, but we add, in quadrature, the full size of the correction to the uncertainty to obtain the final value of $\omega_0=2\pi\times459.1614(28)\,\mathrm{THz}$.  Note that, near to $\omega_0$, the 614\,nm transition has a significant contribution to the curvature of both $\Delta\alpha_0(\omega)$ and $\alpha_2(\omega)$, but these contributions tend to cancel in the ratio.  This explains the excellent linearity in Fig.~\ref{Fig2} and the correspondingly small bias correction.

The measured $\omega_0$ is in excellent agreement with the theoretical value of 459.10(92)\,THz given in \cite{barrett2019polarizability}.  The theory used more accurate, experimentally determined values of $\langle{6 P_{1/2}}\|r\|{6 S_{1/2}}\rangle$ and $\langle{6 P_{3/2}}\|r\|{6 S_{1/2}}\rangle$ from \cite{woods2010dipole}, but $\langle{6 P_{3/2}}\|r\|{5 D_{5/2}}\rangle$ and all uv contributions were determined from atomic structure calculations.  Thus, confirmation of the theoretically estimated zero crossing strongly supports the accurate theoretical assessment of the uv contributions.

As discussed in \cite{barrett2019polarizability}, the zero crossing of $\Delta\alpha_0(\omega)$, 455- and 493-nm transition matrix elements, and branching ratios for $6P_{3/2}$ decays can be combined to provide an accurate model of $\Delta\alpha_0(\omega)$ over a wide wavelength range.  Accurate matrix elements have already been reported \cite{woods2010dipole}, and we have recently carried out improved measurements of $6P_{3/2}$ branching ratios \cite{arnold2020branching}.  Thus the measurement here now allows us to make an accurate assessment of the room temperature blackbody radiation shift for the Ba$^+$ clock transition.  More generally, we give a parametrized model for $\Delta\alpha_0(\omega)$ for wavelengths above 700\,nm. 

Applying the formalism in \cite{barrett2019polarizability}, we approximate $\Delta \alpha_0(\omega)$ by
\begin{equation}
\Delta \alpha_0(\omega)=\Delta\alpha_0^\mathrm{vis}(\omega)-\frac{\Delta\alpha_0^\mathrm{vis}(\omega_0)\left[1-(\omega_0/\omega_a)^2\right]}{1-(\omega/\omega_a)^2}
\end{equation}
where $\omega_a=0.2049(42)\,\mathrm{a.u.}$ is the position of an effective pole that approximates uv contributions and valence-core corrections \cite{barrett2019polarizability}, and $\Delta\alpha_0^\mathrm{vis}(\omega)$ is the contribution from the three transitions at 455, 493, and 614\,nm.  Explicitly,
\begin{multline}
\Delta\alpha_0^\mathrm{vis}(\omega)=\frac{1}{9}\frac{\mu_{455}^2}{\omega_{614}}\frac{1}{1-(\omega/\omega_{614})^2}\left(\frac{\omega_{455}}{\omega_{614}}\right)^3\frac{1-p}{p}\\
-\frac{1}{3}\frac{\mu_{455}^2}{\omega_{455}}\frac{1}{1-(\omega/\omega_{455})^2}-\frac{1}{3}\frac{\mu_{493}^2}{\omega_{493}}\frac{1}{1-(\omega/\omega_{493})^2},
\end{multline}
where $\omega_{614}, \omega_{455},$ and $\omega_{493}$ are the frequencies of the associated transitions expressed in atomic units.  The parameters $\mu_{455}$ and $\mu_{493}$ are the reduced dipole matrix elements
\begin{align}
\mu_{455}&=\langle{6 P_{3/2}}\|r\|{6 S_{1/2}}\rangle=4.7017(27)\,\mathrm{a.u.}\\
\mu_{493}&=\langle{6 P_{1/2}}\|r\|{6 S_{1/2}}\rangle=3.3251(21)\,\mathrm{a.u.}
\end{align}
reported in \cite{woods2010dipole}, and 
\begin{equation}
p=0.763104(65)
\end{equation}
is from recent branching ratio measurements \cite{arnold2020branching}.  From the model we get $\Delta\alpha_0(0)=-73.56(21)\,\mathrm{a.u.}$, consistent with $-73.1(1.3)\,\mathrm{a.u.}$ from theory.  Similarly, $\Delta\alpha_0(\omega_{1762})=-78.51(21)\,\mathrm{a.u.}$, where $\omega_{1762}$ is the clock transition frequency.  For both values given, uncertainties from $\mu_{455}$ and $\mu_{493}$ are assumed correlated and added absolutely, with the result added in quadrature with the uncertainty contributions from $p$, $\omega_0$, and $\omega_a$.  A summary of the uncertainty contributions for $\Delta\alpha_0(0)$ is given in table~\ref{BBR} and those for $\Delta\alpha_0(\omega_{1762})$ are very similar. 
\begin{table}
\caption{Uncertainty contributions to $\Delta\alpha_0(0)$.  Contributions from $\mu_{455}$ and $\mu_{493}$ are added together with the result added in quadrature to the rest.  All uncertainties are the usual 1$\sigma$ values.}
\label{BBR}
\begin{ruledtabular}
\begin{tabular}{l c}
Contribution &  Uncertainty \\
 \hline
 \vspace{-0.2cm}\\
Branching ratio, $p$ & 0.06  \\
455-nm matrix element, $\mu_{455}$ & 0.13 \\
493-nm matrix element, $\mu_{493}$ & 0.05 \\
Zero crossing, $\omega_0$ & 0.02  \\
Effective uv pole placement, $\omega_a$ & 0.08 \\
 \vspace{-0.2cm}\\
Total & 0.21 
 \end{tabular}
 \end{ruledtabular}
 \end{table}

The blackbody radiation (BBR) shift of a clock transition can be written in the form
\begin{equation}
h\delta \nu=-\frac{\Delta\alpha_0(0)E_\mathrm{bbr}^2}{2}\left(\frac{T}{T_0}\right)^4 \left(1+\eta \left(\frac{T}{T_0}\right)^2\right),
\end{equation}
where $T_0=300\,\mathrm{K}$, $E_\mathrm{bbr}=831.945\,\mathrm{Vm^{-1}}$, and $\eta$ is a dynamic correction factor \cite{porsev2006multipolar}.  From the model we estimate the fractional BBR shift at 300\,K to be $3.724(11)\times 10^{-15}$ with a dynamic correction factor $\eta=1.6\times 10^{-3}$.  Thus $\eta$ contributes less than the uncertainty due to the determination of $\Delta\alpha_0(0)$.  We note that all uncertainties quoted are 1$\sigma$.  As discussed in \cite{arnold2020branching}, we would recommend taking a 2$\sigma$ uncertainty in the assessment of clock performance for example.

In summary we have provided an accurate determination of the zero crossing in the dynamic differential scalar polarizability of the $S_{1/2}-D_{5/2}$ clock transition in $^{138}$Ba$^+$.  Measurements utilized a ratio of scalar to tensor components of the differential polarizability so as to eliminate intensity variations in the Stark-shifting laser.  The value of $459.1614(28)\,\mathrm{THz}$ is in excellent agreement with theory.  Following the proposal in \cite{barrett2019polarizability} we have provided a representation of $\Delta\alpha_0(\omega)$ that is accurate at the $0.5\%$ level for wavelengths above 700\,nm.  In principle the model also applies for shorter wavelengths but care would be needed in the vicinity of the zero crossing.  The model will allow accurate calibration of polarizabilities for other clock transitions via comparison with shifts induced on the clock transition in Ba$^+$.

\begin{acknowledgements}
This work is supported by the National Research Foundation, Prime Ministers Office, Singapore and the Ministry of Education, Singapore under the Research Centres of Excellence programme. M. S. Safronova acknowledges the sponsorship of the Office of Naval Research, USA, Grant No. N00014-17-1-2252.
\end{acknowledgements}

\bibliography{BaZero}
\bibliographystyle{unsrt}
\end{document}